\documentclass[english,aps,prb,twocolumn,showpacs,superscriptaddress,10pt]{revtex4-1}

\usepackage[T1]{fontenc}
\usepackage[latin9]{inputenc}
\usepackage{amsmath}
\usepackage{xcolor}
\usepackage{amssymb}
\usepackage{graphicx}
\usepackage{subfigure}
\usepackage{dsfont}
\usepackage[colorlinks,bookmarksopen=true,bookmarksopenlevel=2,bookmarksnumbered=true]{hyperref}

\makeatletter
\providecommand{\tabularnewline}{\\}

\DeclareMathOperator{\sgn}{sgn}

\makeatother

\newcommand{\vc}[1]{\mathbf{#1}}  
\newcommand{\mat}[1]{\mathsf{#1}} 

\renewcommand{\Re}{\mathrm{Re}}
\renewcommand{\Im}{\mathrm{Im}}

\allowdisplaybreaks[1]

\begin{document}

\title{Feshbach-type resonances for two-particle scattering in graphene}

\author{C. Gaul}
\affiliation{Departamento Física de Materiales, Universidad Complutense de
Madrid, E-28040 Madrid, Spain}
\affiliation{Max Planck Institute for the Physics of Complex Systems, 01187 Dresden, Germany}
\author{F. Domínguez-Adame}
\affiliation{Departamento Física de Materiales, Universidad Complutense de
Madrid, E-28040 Madrid, Spain}
\author{F. Sols}
\affiliation{Departamento Física de Materiales, Universidad Complutense de
Madrid, E-28040 Madrid, Spain}
\author{I. Zapata}
\affiliation{Departamento Física de Materiales, Universidad Complutense de
Madrid, E-28040 Madrid, Spain}
\affiliation{Departamento Física de la Materia Condensada C-3, Universidad Autónoma de Madrid, E-28049 Madrid, Spain}


\begin{abstract}
Two-particle scattering in graphene is a multichannel problem, where the energies of the identical or opposite-helicity channels lie in disjoint energy segments. Due to the absence of Galilean invariance, these segments depend on the total momentum $Q$.  The dispersion relations for the two opposite-helicity scattering channels are analogous to those of two one-dimensional tight-binding lattices with opposite dispersion relations, which are known to easily bind states at their edges.
When an $s$-wave separable interaction potential is assumed, those bound states reveal themselves as three Feshbach resonances in the identical-helicity channel. In the limit $Q \rightarrow 0$, one of the resonances survives and the opposite-helicity scattering amplitudes vanish.
\end{abstract}

\maketitle

\section{Introduction}

Since the advent of graphene,\cite{Novoselov2004} whose low-energy excitations
behave as massless and chiral Dirac fermions propagating in two
dimensions~(2D),\cite{Wallace1947,CastroNeto2009} the role of
elec\-tron-electron interactions has been an active research
field.\cite{kotov2012electron} In this context, the two-body problem in a
single layer of graphene has been studied
recently.\cite{sabio2010two,lee2012quasilocalized}
Reference~\onlinecite{sabio2010two} found that, due to the relativistic
dispersion relation of electrons in graphene, the conventional decoupling of
center-of-momentum and relative coordinates fails, which prevents a simple
effective one-body description. The general case of nonzero total momentum was
barely addressed in Ref.~\onlinecite{sabio2010two} despite its potential
importance for charge transport phenomena in graphene.\cite{Peres2010} The main
goal of our present work is to contribute to fill this gap by carrying out a
detailed analysis of the two-particle scattering problem at nonzero total
momentum. Like the work of Refs.~\onlinecite{sabio2010two,lee2012quasilocalized},
the study presented here could provide important insights on the many-body physics of graphene.

One aspect of the two-body Dirac scattering in 2D which has so far received
little attention is its multi-channel character. A remarkable  feature of
multichannel scattering of particles with internal structure is the occurrence
of Fano-Feshbach resonances,\cite{Fano1935,Fano1961,Feshbach1958} extensively
studied in nuclear and atomic physics and with a wealth of recent applications
to quantum gases.\cite{kohler2006production,chin2010feshbach} In this work we
show that similar resonances appear for the two-body problem in graphene.

\section{Two-body problem in graphene}

We consider the scattering of two interacting particles moving in a perfect
graphene lattice. If their crystal momenta are close enough to the same Dirac
point, a continuum description in a single valley suffices.
Then, the wave function $\Phi_{\alpha \beta}(\mathbf{r}_1,\mathbf{r}_2)$ for two particles has four components,
the double index  $\alpha,\beta$ referring to the sublattice (pseudo-spin) indices of particles 1 and
2 respectively.
At energy $\epsilon$, $\Phi_{\alpha\beta}$ is governed by the Dirac equation (we use units in which $\hbar=v_F=1$)
\begin{equation}\label{EOM_realspace}
\epsilon \Phi_{\alpha\beta}
 =- i \boldsymbol{\sigma}_{\alpha\alpha'} \!\cdot\! \boldsymbol{\nabla}_{\!1} \Phi_{\alpha'\beta}
  - i \boldsymbol{\sigma}_{\beta\beta'}   \!\cdot\! \boldsymbol{\nabla}_{\!2} \Phi_{\alpha\beta'}
  + V \Phi_{\alpha\beta} \ ,
\end{equation}
where doubly appearing indices $\alpha'$ and $\beta'$ are summed over.
The two-dimensional gradient $\boldsymbol{\nabla}_{\!j}$ acts on the coordinate of particle $j$,
$\boldsymbol{\sigma}$ is the vector of Pauli matrices,
and $V=V(\mathbf{r}_1-\mathbf{r}_2)$ is the two-body interaction.

Since Eq.\ \eqref{EOM_realspace} conserves the total momentum $\mathbf{Q}$ (measured with respect to the Dirac point), we choose to write the wave function as follows
\begin{equation}
\Phi_{\alpha \beta}(\mathbf{r}_{1},\mathbf{r}_{2})= e^{i\left[\mathbf{Q}\cdot \left(\mathbf{r}_{1}+\mathbf{r}_{2}\right)/2-\epsilon
t\right]}\Psi_{\alpha\beta}( \mathbf{r}_{1}-\mathbf{r}_{2}) \ .
\end{equation}
As the center-of-mass and relative motions of two particles in a
graphene lattice do not factorize,\cite{sabio2010two} the relative wave
function $\Psi_{\alpha\beta}(\mathbf{r}_{1}-\mathbf{r}_{2})$ depends on the
total momentum $\mathbf{Q}$, which appears as a parameter in the relative
two-body problem.

The Fourier transform of the relative wave function is the four-component vector
$\Psi(\mathbf{q})$, where $\mathbf{q}$ is half the relative momentum, hereafter
expressed in complex notation ($\mathbf{q} \rightarrow q \in \mathbb{C}$ and
$\mathbf{Q} \rightarrow Q \in \mathbb{R}^+$ without loss of generality).
In the helicity representation (see Appendix \ref{sec:AppDiracEqi}), the Dirac equation \eqref{EOM_realspace} reads
\begin{equation}
\epsilon\Psi(q)  = \mat{K}(q) \Psi(q) +
\frac{1}{\Omega}\sum_{q'}V(q,q')\,\mat{U}^{\dagger}(q)\mat{U}(q')\Psi(q')\ ,
\label{eq:Dirac2BHelicity}
\end{equation}
where $\Omega$ is the area. The $4 \times 4$ matrices $\mat{K}(q)$ and
$\mat{U}(q)$ are implicit functions of the total momentum $Q$. We choose to work
in the valley where energy and helicity have the same sign. The kinetic energy $\mat{K}(q)$ is given by the diagonal matrix
\begin{equation}
\newcommand{\hsp}{\hspace*{-2.5ex}}
\mat{K}(q)=\left[\begin{array}{ccccc}
|q_{+}|\!+\!|q_{-}| \\
        &\hsp\phantom{-}|q_{+}|\!-\!|q_{-}|\\
        &       &\hsp-|q_{+}|\!+\!|q_{-}|\\
        &       &        &\hsp-|q_{+}|\!-\!|q_{-}|
\end{array}\right],\label{eq:KineticEnergy2B}
\end{equation}
with $q_{\pm}=Q/2\pm q$ the momentum of the electrons, where we use units
$\hbar=v_F=1$. The unitary matrix $\mat{U}(q)=\mat{U}_{d}(q)\mat{R}^\dagger$ is
related to the transformation from sublattice coordinates to helicities and consists in
\begin{eqnarray}
\mat{U}_{d}(q)&=&\exp(-i\theta_{+}\sigma^{3}/2)\otimes
\exp(-i\theta_{-}\sigma^{3}/2) \nonumber \\
\mat{R}&=& \exp(-i\pi\sigma^{2}/4)\otimes \exp(-i\pi\sigma^{2}/4),
\end{eqnarray}
$\sigma^{i}$ being the Pauli matrices and $\theta_{\pm}=\arg(q_{\pm})$.
Finally, $V(q,q')$ is the interaction matrix element for the two-particle
scattering process. It is independent of $Q$ and the helicity indices.

The kinetic energy $\epsilon$ lies in three non-overlapping intervals for the
four possible helicity channels: $(-\infty,-Q)$ for channel $\left(--\right)$,
$(-Q,Q)$ for channels $\left(+-\right),\left(-+\right)$, and $(Q,\infty)$ for
channel $\left(++\right)$. Thus, the elastic collision of two electrons in
graphene poses a multichannel problem, with the important peculiarity that,
unlike in atomic multichannel scattering, open channels close when energy
crosses thresholds. We note that, in the isotropic limit, $Q\rightarrow0$, the
central energy interval collapses to a point at $\epsilon=0$, which results in a
nontrivial scattering problem.

Because the two opposite-helicity channels have a bounded energy range, we
expect that resonances may appear in the $\left(--\right)$ or $\left(++\right)$
channels ($|\epsilon|>Q$). These resonances are due to virtual transitions into
quasi-bound states of the $\left(+-\right),\left(-+\right)$ channels
($|\epsilon|<Q$), which would be true bound states in the absence of coupling
between channels. This is the same mechanism which underlies Fano-Feshbach
resonances in atomic and nuclear physics. What is unique to the resonances
encountered here is the absence of particle internal structure and the
fundamental sensitivity to the absolute motion.

\subsection{Symmetries}

Negative and positive energies are related by a simple symmetry operation.
Consider the matrix $\mat{m}_1=\sigma^1\otimes\sigma^1$, which interchanges the
helicities of both particles. It commutes with $\mat{U}^\dagger(q)\mat{U}(q')$
and anticommutes  with $\mat{K}(q)$ in Eq.~\eqref{eq:Dirac2BHelicity}. Thus,
applying the transformation $\mat{m}_1$ and changing the sign of the interaction
$V(q,q')$ is equivalent to changing the sign of the energy. Thus, we may
restrict ourselves to positive energies in the rest of this work.

The present problem lacks parity,  rotation  and time-reversal symmetry.
However, we identify two relevant symmetries:

(i) Permutation of the two colliding particles $\mat{P}_{12}$. This exchange
operator can be written as $\mat{P}_{12}=\hat{P}\mat{m}_{12}$, where
$\hat{P} q \hat{P} = -q$, and $\mat{m}_{12}$
interchanges the pseudo-spin components and (because $[\mat{m}_{12},\mat{R}]=0$)
helicities $\left(+-\right)\leftrightarrow\left(-+\right)$. It is easy to prove
that $[\mat{P}_{12},\mat{K}]=[\mat{P}_{12},\mat{U}]=0$. Therefore, if the
interaction has the symmetry $V(q,q')=V(-q,-q')$, then
$\mat{P}_{12}$ commutes with the Hamiltonian.

(ii) Reflection at the $x$-axis  $\mat{P}_{x}=\hat{P}_x \mat{m}_3$, where
$\hat{P}_x q \hat{P}_x = q^*$, and $\mat{m}_{3}= \sigma^3 \otimes \sigma^3$.
If $V(q,q')=V(q^*,{q'}^*)$, then $\mat{P}_x$ commutes with the Hamiltonian.

We will classify the scattering states according to these symmetries later.

\subsection{\texorpdfstring{$T$}{T}-matrix equation and solution}

We consider a purely $s$-wave separable potential $V(q,q')=\lambda_0$ hereafter,
such that both symmetries are fulfilled. We will find that even in this simple
case, the scattering amplitude displays a rich structure. The $T$-matrix
equation for the two-body scattering problem formulated in
Eq.~\eqref{eq:Dirac2BHelicity} satisfies
\begin{align}
\mat{T}(z;q_f,q_i) & = \mat{W}(q_f,q_i) \nonumber \\*
&+\frac{1}{\Omega}\sum_{q}\mat{W}(q_f,q)\mat{G}_{0}(z;q)\mat{T}(z;q,q_i)\ ,
\label{eq:TMatrix2BEq}
\end{align}
where  $\mat{W}(q,q') = \lambda_{0} \mat{U}^{\dagger}(q)\mat{U}(q')$
incorporates interaction and pseudo-spin rotation, and $\mat{G}_{0}(z;q)=
\left(z-\mat{K} \right)^{-1}$ is the unperturbed propagator. Introducing an
upper cutoff $p_c$, the solution of Eq.~\eqref{eq:TMatrix2BEq} is
\begin{equation}
\mat{T}(z;q_f,q_i) = \mat{U}^{\dagger}(q_f)\left[\lambda_{0}^{-1}-
\mat{M}(z)\right]^{-1}\mat{U}(q_i)\ ,
\label{eq:TMatrix2BSol1}
\end{equation}
with
$$
\mat{M}(z) = \frac{1}{4\pi^{2}}\int_{|q|<p_{c}}\!\!d^{2}q\,
\mat{U}(q)\mat{G}_{0}(z;q)\mat{U}^{\dagger}(q)\ .
$$
Note that, thanks to the separable $s$-wave potential, this solution is
\emph{exact}.

\section{Results}

In the $q$-plane, the curves of constant kinetic energy are either homo-focal
ellipses [channels $(++)$ and $(--)$, for $|\epsilon|>Q$] or homo-focal
hyperbolae [channels $(+-)$ and $(-+)$, for $|\epsilon|<Q$], which cross at
right angles. This suggests the use of elliptic coordinates (see
Appendix~\ref{sec:AppElliptic-coordinates}).  Specifically, the transformation
$q=(Q/2)\cosh\left(u+iv\right)$, with $u\geq0$ and $-\pi<v\leq\pi$, renders the
kinetic energy separable,
\begin{equation}
\mat{K}(q)=Q\left[\begin{array}{cccc}
\cosh u \\
& \cos v \\
& & -\cos v \\
& & & -\cosh u
\end{array}\right].
\end{equation}

The $v$-dependence of the kinetic energy in the two central channels
$\left(+-\right),\left(-+\right)$ resembles the dispersion relation of a Bloch
wave in a tight-binding chain with nearest-neighbor hopping
and $v$ playing the role of crystal momentum. In this picture, we would expect
the interaction $V$ to play the role of an impurity potential that can nucleate
bound states lying outside the band $(-Q,Q)$, where they become resonances for
the outer ($--,++$) channels with incoming energy $|\epsilon|>Q$.
Mathematically, this translates into the appearance of poles for the
outer-channel propagators in the lower part of the complex-energy plane, and
hence in the $z$-dependence of the $T$-matrix~\eqref{eq:TMatrix2BSol1}.

The matrix $\mat{M}(z)$ in Eq.~\eqref{eq:TMatrix2BSol1}  can be computed by
introducing elliptic coordinates. It becomes of the form
\begin{equation}
\mat{M}=\left[
  \begin{array}{ll}
    \mat{A}        & \mat{B}\\
    \mat{B}^{\intercal} & \mat{A}
  \end{array}\right] \ ,
\quad
\mat{A}=\left[
  \begin{array}{cc}
    d & a\\
    a & d
  \end{array}\right] \ ,
\quad
\mat{B}=\left[
  \begin{array}{cc}
    a & b\\
    c & a
  \end{array}\right] \ ,
\label{eq:MPrimeSolution1}
\end{equation}
where $a,b,c,d$ are complex functions of $z$, $Q$, and $p_c$
and are given in Appendix~\ref{sec:AppElliptic-coordinates}. The matrix
$\mat{M}(z)$ inherits the symmetries $\mat{P}_{12}$ and $\mat{P}_{x}$ of the
scattering problem Eq.~\eqref{eq:Dirac2BHelicity}: $\mat{M}(z)$ commutes with
both $\mat{m}_{12}$ and $\mat{m}_{1}=\mat{R}^{\dagger} \mat{m}_{3}
\mat{R}=\sigma^1 \otimes \sigma^1$. The eigenvectors can be classified as
symmetric and antisymmetric under $\mat{m}_{12}$ and $\mat{m}_{1}$, which
defines corresponding eigenspaces of the $T$-matrix Eq.~\eqref{eq:TMatrix2BSol1}
(with respect to $\mat{P}_{12}$ and $\mat{P}_{x}$). In the antisymmetric
eigenspace of $\mat{m}_1$, two eigenvectors
$v_{a}^\intercal=(0,1,-1,0)/\sqrt{2}$ and
$v_{s}^{\intercal}=(1,0,0,-1)/\sqrt{2}$ of $\mat{M}(z)$ with eigenvalues
$d(z)-c(z)$ and $d(z)-b(z)$ are found which are antisymmetric and symmetric
under $\mat{m}_{12}$, respectively.
In the symmetric eigenspace of $\mat{m}_1$, two eigenvectors of the form
$v_{s\pm}^{\intercal}(z) = [a_{\pm}(z),\pm b_{\pm}(z),\pm
b_{\pm}(z),a_{\pm}(z)]$ are found, both being also symmetric under
$\mat{m}_{12}$.
All eigenvectors are normalized as $v_{i}^{\intercal}(z)v_{j}(z)=\delta_{ij}$.
The explicit formulae for $a_\pm / b_\pm$ are given in
Appendix~\ref{sec:AppElliptic-coordinates}. In the basis of eigenstates of
$\mat{m}_{12}$ and $\mat{m}_{1}$, we can write
\begin{align}
\bigl[\lambda_{0}^{-1} -\mat{M}(z)\bigr]^{-1}
&= \hspace{-1ex}\sum_{j=a,s,s\pm} \hspace{-1ex}
     t_{j}(z)v_{j}(z)v_{j}^{\intercal}(z) \ .
\label{eq:MPrimeSpectralDecomp}
\end{align}

The eigenvalues of the $T$-matrix, $t_j(z)$, are functions of $a,b,c,d$ and
can be calculated to be
\begin{subequations}
\label{eq:tMatValuesX}
\begin{align}
\frac{1}{t_{a}(z)} & = \frac{1}{\lambda_{0}} + \frac{Q}{8\pi}h(x)g(x)\ ,
\label{7a}\\
\frac{1}{t_{s}(z)} & = \frac{1}{\lambda_{0}} +
\frac{Q}{4\pi}\left[\Gamma_{c}^{2}\rho(x)+\frac{1}{4}\,h(x)g(x)\right]\ , \\
\frac{1}{t_{s\pm}(z)} & = \frac{1}{\lambda_{0}} +
\frac{Q}{8\pi}\biggl[\Gamma_{c}^{2}\,
\frac{\rho(x)\left(x\mp\sqrt{x^{2}+3}\,\right)-1}{h(x)} \nonumber\\*
& + \frac{1}{4}\,h_{\pm}(x)g(x)\biggr]\ , \label{7c}
\end{align}
\end{subequations}
where $x=z/Q$,
$h(x) = \sqrt{x^{2}-1}$,
$g(x) = \ln (4\Gamma_{c})- \ln \left[x+h(x)\right]
      + i(\pi/2)\sgn\left[\Im(x)\right]$,
$\Gamma_{c} = p_{c}/Q$,
$\rho(x) = h(x) -x$, and finally
$$
h_{\pm}(x) =
 \frac{1}{h(x)}\left[x^{2}+1 \pm
  \frac{h(x)+x\big[3 + x(2+x^{2})\rho(x)\big]}{\sqrt{x^{4}+2x^{2}-3}\,\rho(x)}
  \right]  .
$$
For $|x|<1$, the usual analytical continuation $\sqrt{x^{2}-1} \to
i\sqrt{1-x^{2}}$ is implied. For $t_{s\pm}(z)$, we have shown only the leading
behavior in the cutoff, which is sufficient for the low-energy region considered
here.

To compute the scattering amplitudes, we must consider on-shell expressions,
thus letting $z=\epsilon+i0^{+}$ where the dependence of $\epsilon$ on the
initial and final momenta is channel specific. For example, for $(++)$
scattering one has $\epsilon=|q^{i}_{+}|+|q^{i}_{-}|=|q^{f}_{+}|+|q^{f}_{-}|$.

\subsection{Non-resonant scattering}

Due to the quadratic dependence on the cutoff of $t^{-1}_{s}$ and
$t^{-1}_{s\pm}$, for most energies within the relevant region $\epsilon\ll
p_{c}$ the collision will be dominated by the antisymmetric (in $\mat{P}_{12}$)
element alone, $|t_a| \gg |t_{s\pm}|,|t_s|$. By using Eq.~\eqref{7a} and the
projector element from Eq.~\eqref{eq:aProjector},
we can derive the final formula
for the scattering amplitude for two particles colliding in the antisymmetric
mode of the $(++)$ incoming channel
\begin{align}
\label{non-resonant}
& T_{++,++}\left(\epsilon+i0^{+}; q_f,q_i \right) \nonumber \\*
&=\frac{1}{2}
\frac{
\sin\bigl[({\theta^{f}_{-}-\theta^{f}_{+}})/{2}\bigr]
\sin\bigl[({\theta^{i}_{-}-\theta^{i}_{+}})/{2}\bigr]
}
{\lambda_{0}^{-1}+\frac{1}{8\pi}\sqrt{\epsilon^{2}-Q^{2}}
\left(\ln\frac{4p_{c}}{\epsilon+
\sqrt{\epsilon^{2}-Q^{2}}}+i\frac{\pi}{2}\right)}\ .
\end{align}

\subsection{Feshbach resonances}\label{subsecFeshbachResonances}

As can be guessed from Eq.~\eqref{eq:tMatValuesX}, the symmetric diagonal
$T$-matrix elements are relevant only when the real part of their inverses is
very small, $\Re\bigl(t_{j}(z)^{-1}\bigr)\simeq 0$. However, for the
$\left(+-\right)$ and $\left(-+\right)$ channels ($|x|<1$), because in that case
$\Im\bigl(t_{j}(z)^{-1}\bigr)$ scales with $p_c^2$ and thus is always large, the
symmetric $T$-matrix element will be very small. By contrast, for the two
($|x|>1$) channels [$\left(++\right)$ if $\lambda_{0}>0$ and $\left(--\right)$
if $\lambda_{0}<0$], the possibility of a sharp resonance exists since $\Im
\left(g(z)\right)$ is small and independent of $p_c$.

It follows from Eq.~\eqref{eq:tMatValuesX} that, by just taking into account the
dominant quadratic dependence on the cutoff, the equation
$\Re\left(t_{s}(z_s)^{-1}\right)= 0$ can be satisfied at $x_{s}\equiv
\epsilon_s/Q$ if
\begin{subequations}
\begin{equation}
\lambda_{0}p_{c}^{2}/4\pi Q = -1/\rho(x_{s})\ .
\label{eq:Resonance_s}
\end{equation}

Similarly, in the case of $t_{s\pm}(z)$ given by Eq.~\eqref{7c}, the location of
the resonance is given by
\begin{equation}
\frac{\lambda_{0} p_{c}^{2}}{4\pi Q} =
\frac{2\sqrt{x_{s\pm}^{2}-1}}{1-\rho(x_{s\pm})\left(x_{s\pm} \mp
\sqrt{3+x_{s\pm}^{2}}\,\right)}\ .
\label{eq:Resonance_spm}
\end{equation}
\label{all_resonances}
\end{subequations}

\begin{figure}[t]
\includegraphics[width=0.85\linewidth,clip,trim=0 0 0 0]{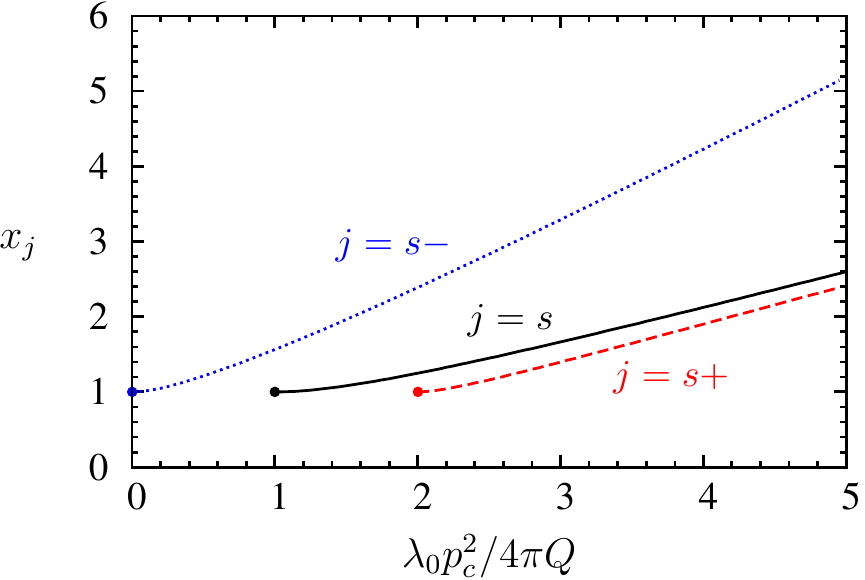}
\caption{Location of the (asymmetric) resonances Eq.~\eqref{eq:tMatValuesX},
defined by the condition $\Re[t_{j}(x_{j} Q)^{-1}] = 0$.}
\label{fig:Resonance_Location}
\end{figure}

\begin{figure}[t]
\includegraphics[width=0.9\linewidth]{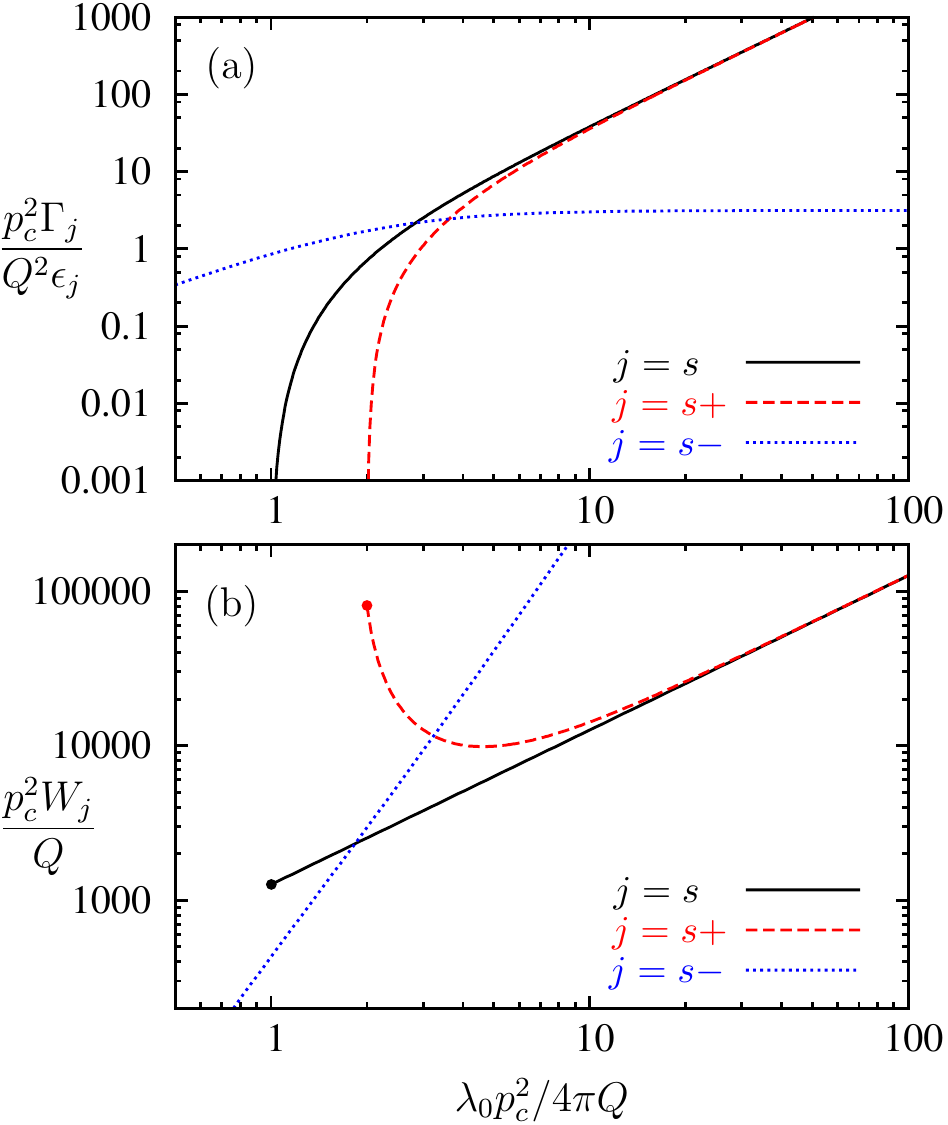}
\caption{Width $\Gamma_j$ and weight $W_j$ of the symmetric resonances, defined
via 
$2\pi |t_j(z)|^2 = W_j     \Gamma_j/[(z-\epsilon_j)^2 + \Gamma_j^2/4]$
repulsive interaction $\lambda_0>0$.  The relative width $\Gamma_j/\epsilon_j$
shown in (a) is measured in units of $\Gamma_c^{-2} \ll 1$, i.e., the resonances
are always narrow.}
\label{fig:Resonance_Width-Weight}
\end{figure}

As shown in Fig.~\ref{fig:Resonance_Location}, the solutions $x_j$ of
Eqs.~\eqref{all_resonances} are monotonically growing, positive functions of
${\lambda_{0} p_{c}^{2}}/{4\pi Q}$ that start at 0, 1, 2, for $j=s-$, $s$, and
$s+$, respectively. Consequently, the interaction strength ${\lambda_{0}
p_{c}^{2}}/{4\pi Q}$ must equal or exceed the thresholds $c_{s-}=0$, $c_s=1$,
and $c_{s+}=2$ for the corresponding resonances to occur. The asymptotics for
$x_j\gg 1$ are
\begin{equation}
x_j \simeq b_j \frac{\lambda_0 p_c^2}{4\pi Q}\ ,
\qquad
b_{s-}=1\ ,
\quad
b_s=b_{s+}=\frac{1}{2}\ .
\end{equation}
Thus, for the resonance to occur and the resulting resonance energy to lie in
the region where the approximations are valid ($\epsilon_{j}\ll p_{c}$), the
following condition is necessary and sufficient
\begin{equation}
c_j b_j \frac{Q}{p_{c}}<b_j\frac{\lambda_{0}p_{c}}{4\pi}\ll 1\ .
\label{eq:ts1ResonanceConditions}
\end{equation}
In this situation the inclusion of the other sub-dominant real part only shifts
the position of the resonance by a negligible amount. By Taylor expanding
$t_{j}(z)^{-1}$ near the resonance, one arrives at a Breit-Wigner form
\begin{equation}
t_{j}(z) \approx
\sqrt{ \frac{W_j \Gamma_j}{2\pi}}
\frac{1}{z-\epsilon_{j} + i\Gamma_{j}/2}\ ,\quad j=s,s+,s- \ ,
\end{equation}
with weight $W_j$ and width $\Gamma_j$. For $j=s$, the results are rather
simple, $\Gamma_{s} \simeq  \lambda_{0}\left(\epsilon_{s}^{2}-Q^{2}\right)/16$
and $W_s \simeq 32\pi\lambda_{0}$. The ratio
$\Gamma_{s}/\epsilon_{s}<\lambda_{0}\epsilon_{s}/16\ll1$ is very small, implying
a narrow resonance. For $j=s\pm$, the formulas are lengthy. From their width
shown in Fig.~\ref{fig:Resonance_Width-Weight}(a) (along that of $j=s$) one
concludes that the Breit-Wigner form survives; hence they are narrow as well.
The weight of all resonances are shown in
Fig.~\ref{fig:Resonance_Width-Weight}(b).

The contribution of the resonance $s$ to the scattering amplitude near the
resonance reads
\begin{align}\label{eq:resonantT_s}
&T_{++,++}\left(\epsilon+i0^{+};q_f,q_i\right) \nonumber \\*
&\simeq \frac{1}{2} \, t_{s}(\epsilon)
\sin\bigl[({\theta^{f}_{-}+\theta^{f}_{+}})/{2}\bigr]
\sin\bigl[({\theta^{i}_{-}+\theta^{i}_{+}})/{2}\bigr]\ .
\end{align}
The expressions for the other two resonances $s\pm$ are much more involved.

The fact that these resonances are of the Fano-Feshbach type is confirmed by a
simple computation in which one neglects the coupling between channels. Then,
precisely at the energy given in  Eq.~\eqref{eq:Resonance_s}, one finds a bound
state for the symmetric $\left(+-\right)$ and $\left(-+\right)$ channel. Once
the inter-channel coupling is included, this bound state turns into a sharp
resonance. This is exactly the same mechanism underlying the appearance of
Fano-Feshbach resonances. Close to the resonance energy, the incoming particle
pair, say, in the symmetric subspace of the $\left(++\right)$ channel with
$\lambda_{0}>0$ virtually jumps into the quasi-bound state in the symmetric
subspace of the $\left(+-\right)$ and $\left(-+\right)$ channels. However,
unlike in atomic Fano-Feshbach resonances, the neglect of inter-channel coupling
is not a good approximation in the present problem mostly because of the
locality of the interaction, which strongly mixes the channels. This explains
why the other two resonances are not obtained from the same type of reasoning.

The present approach makes a poor prediction about the resonant energies because
of their strong dependence on the cutoff. In fact, for this model $\lambda_0
p_c^2/4\pi=\pi V_0$, where $V_0$ is the interaction at the origin in real space.
We can restore units to show that the dimensionless parameter which controls all
the characteristics of the resonances is nothing but the ratio of energies $\pi
V_0/\hbar v_F Q$. On the other hand, the relative characteristics of the
resonances can be shown to be insensitive to variations in the high-momentum
content of the interaction if restricted to be of the form $V(q,q')=\lambda_0
f(|q|)f^*(|q'|)$. Specifically, it can be proven that
Eqs.~\eqref{all_resonances} remain valid provided that $p_c^2$ is replaced by a
single function of an effective momentum cutoff. This means that for this
restricted set of interactions, once a resonance is given, the properties of the
other two are independent not only of the cutoff but also of the other model
parameters. One could object that total momentum is not strictly conserved for
separable interactions. However, momentum conservation does not have to be
conserved beyond the Dirac approximation, when umklapp processes in the electron
interaction as well as finite bandwidth effects are taken into account.

\subsection{Isotropic limit}

The isotropic limit for the $(++)$ channel can be obtained immediately from the
previous formulae. One has to remember that in this limit,
$\theta_{+}-\theta_{-}\rightarrow\pi$, and we take $\epsilon\gg Q$. Some
degeneracy of the $T$-matrix must be taken into account but the calculation is
otherwise straightforward. We give here the result for the scattering amplitude
of two particles colliding in the $(++)$ channel
\begin{align}
&T_{++,++}\left(\epsilon+i0^+; q_f,q_i \right) = \frac{1}{2}\,
\left[ \frac{1}{\lambda_0^{-1}+
\frac{\epsilon}{8\pi}\left( \ln \frac{2p_c}{\epsilon}
+ i \frac{\pi}{2} \right)} \right.
\nonumber \\*
&+ \left. \frac{\cos \left( \alpha_f-\alpha_i \right)}%
{\lambda_0^{-1}-\frac{p_c^2}{8\pi \epsilon} +
\frac{z}{16\pi}\left( \ln \frac{2p_c}{\epsilon}
+ i \frac{\pi}{2} \right)} \right]\, ,
\end{align}
where $\alpha_{i,f} \equiv {\rm arg}(q_{i,f})$.

The first term is similar to that encountered in the one-body scattering by an
impurity (see Appendix \ref{sec:AppImpScatt}), with a smooth behavior as a
function of energy.  By contrast, the second term is specific of two-body
scattering and displays a sharp resonance provided the second condition in
Eq.~\eqref{eq:ts1ResonanceConditions} is fulfilled. The well defined limits for
the position and the width of the resonance are
$
\epsilon_{r}  \simeq \lambda_{0}p_{c}^{2}/8\pi
$
and
$
\Gamma  \simeq \lambda_{0}\epsilon_{r}^2/16
$.

On the other hand, the isotropic limit in the $(+-)$-$(-+)$ channels is more
delicate. As the kinetic energy range collapses to a point, we have to use a
different approach. Here we fix the incoming momenta and let $Q\to 0$. This
implies that the energy has to vanish as well, with $\epsilon \sim Q \cos
\alpha_{i}$ and $q_i=|q_{i}| \exp(i\alpha_{i})$. This implies that $\cos
\alpha_{f} = \pm \cos \alpha_{i}$ because of the possible helicity flip, but
there is no restriction on $|q_{f}|$.

The leading behavior for the scattering amplitudes in these channels is $t_a
(\epsilon) \sim Q^2$ while $t_s (\epsilon)$ and $t_{s\pm} (\epsilon)$ scale as
$Q/p_c^2$, which clearly shows that scattering amplitudes vanish in this
isotropic limit (see Appendix~\ref{sec:AppIsotrLimit}). A possible explanation
for this fact relies again on the analogy with the tight-binding Hamiltonian.
The $Q\to 0$ limit can be seen as the limit in which the width of the band
becomes very narrow. The effective mass of the tight-binding particles increases
without limit, meaning that any state prepared with a given lattice momentum
$q_i$ will suffer very little scattering.

\section{Conclusions}

We have identified an analogy between the two-particle scattering in graphene
and the Fano-Feshbach effect in atomic and nuclear physics. As an example, the
case of a $s$-wave separable potential has been fully analyzed. A set of
resonances has been found whose origin is traced back to virtual transitions
into closed channels. Our study establishes a connection between the fields of
electronic interactions in graphene and Fano-Feshbach resonances in cold atom
systems.

\begin{acknowledgments}

The authors thank F. Guinea and N. Zinner for helpful comments. This work was
supported by MINECO through grants FIS2010-21372 and  MAT2010-17180, by
Comunidad de Madrid through grant Microseres-CM, and by the EU through
Marie Curie ITN NanoCTM. Research of C.G.\ was supported
by a PICATA postdoctoral fellowship from the Moncloa Campus of International
Excellence (UCM-UPM).

\end{acknowledgments}

\appendix

\section{Dirac equation in graphene\label{sec:AppDiracEqi}}

\paragraph{Free Dirac equation.} We use units in which $\hbar=v_{F}=1$ and the
Einstein summation convention on repeated indexes, unless the repeated index
appear at both sides of the equation, being orphaned in only one of the sides.
Indexes given by the first letters of the Greek alphabet, $\alpha,\beta,\ldots =
\,\uparrow,\downarrow$, refer to pseudo-spin (sub-lattice quantum number),
those with Latin letters, $j,k=1,2\equiv x,y$, will run in 2D-Euclidean space
and those with late letters of the Greek alphabet, $\sigma,\tau,\ldots=+,-$,
refer to the helicity of the particle. The single-valley one-particle Dirac
equation in graphene in real space reads
\begin{equation}
i\,\partial_{t}\psi_{\alpha}(\vc{r},t)=-i\,\partial_{j}\sigma_{\alpha\beta}^{j}
\psi_{\beta}(\vc{r},t)+V(\vc{r})\psi_{\alpha}(\vc{r},t)\ ,
\label{eq:DiracRealSpace}
\end{equation}
where $\vc{r}=(x,y)$ and $\sigma^{j},\, j=1,2,3$ denote the Pauli matrices.

When $V(\vc{r})=0$, the plane wave solutions can be chosen as
$\psi_{\alpha}(\vc{r},t)\sim \mat{w}_{\alpha\sigma}(\vc{k})\, e^{i\left(\vc{k}
\cdot \vc{r}-\epsilon t\right)}$, where $\vc{k}$ is the momentum and
$\epsilon=\sigma|\vc{k}|$ is the energy. The columns of $\mat{w}(\vc{k})$ are
normalized spinors and satisfy
\begin{equation}
k_{j}\sigma_{\alpha\beta}^{j}\mat{w}_{\beta\sigma}(\vc{k})=
\sigma k\, \mat{w}_{\alpha\sigma}(\vc{k}).
\label{eq:DiracMomentumSpace}
\end{equation}
Orthonormality and completeness for the spinors read
$
\mat{w}^{\dagger}(\vc{k})\, \mat{w}(\vc{k}) = \mat{w}(\vc{k})\,
\mat{w}^{\dagger}(\vc{k}) = \mat{1}
$.
Specifically, we chose the spinors as in Ref.~\onlinecite{CastroNeto2009}:
\begin{equation}
\mat{w}(\vc{k})=\frac{1}{\sqrt{2}}\left(\begin{array}{cc}
e^{-i\theta_{k}/2} 	& e^{-i\theta_{k}/2}\\
e^{i\theta_{k}/2}	&-e^{-i\theta_{k}/2}
\end{array}\right)\ ,
\label{eq:DiracSpinors}
\end{equation}
with $\vc{k}=k\left(\cos \theta_{k},\sin \theta_{k}\right)$.
For later use we collect some useful formulas here
\begin{align*}
 \mat{w}^{\dagger}(\vc{k}_{1})\, \mat{w}(\vc{k_{2}})
 &=  e^{ i (\theta_1 - \theta_2) \sigma^1/2}
  = \mat{u}^{\dagger}(\vc{k}_{1})\, \mat{u}(\vc{k_{2}}) \ ,
  \label{eq:SpinorProp1} \\
 \mat{u}(\vc{k}) = \mat{w}(\vc{k})\sigma^1
 &= e^{-i \theta_k \sigma^3/2} \, e^{i \pi \sigma^2/4}
 = \mat{u}_d(\vc{k}) \, \mat{r}^\dagger\ .
\end{align*}

Normalized solutions of the one-particle Dirac equation will be written in real
and Fourier spaces, and helicity or pseudo-spin basis, as
\begin{align}
\psi_{\alpha}(\vc{k},t) & = \mat{w}_{\alpha \sigma}(\vc{k})\,
\psi_{\sigma}(\vc{k},t)
   =\frac{1}{\sqrt{\Omega}}\int_{\smash{\Omega}}d^{2}r\,
   \psi_{\alpha}(\vc{r},t)\, e^{-i\vc{k}\cdot\vc{r}},\nonumber\\
\psi_{\sigma}(\vc{k},t) & = \mat{w}_{\sigma\alpha}^{\dagger}(\vc{k})\,
\psi_{\alpha}(\vc{k},t),\nonumber\\
\psi_{\alpha}(\vc{r},t) &=\frac{1}{\sqrt{\Omega}}\sum_{k}\,
\psi_{\alpha}(\vc{k},t)\, e^{i\vc{k}\cdot\vc{r}}\ .
\end{align}

\paragraph{Interaction potential.} The potential $V(r)$ is assumed to be
spherically symmetric and short ranged. The Fourier transform reads
\begin{align}
V(\vc{k},\vc{q}) &=\int_{\Omega}d^{2}r\, e^{-i(\vc{k}-\vc{q})\cdot\vc{r}}V(r)
\nonumber\\
&=\sum_{l=-\infty}^{\infty}e^{i l(\theta_{q}-\theta_{k})}V_{l}(k,q)\ ,
\end{align}
with
$$
V_{l}(k,q) = 2\pi\int_{0}^{\infty}dr\, r\, V(r)\, J_{l}(kr)\, J_{l}(qr)\ ,
$$
where $J_{n}(z)$ is the Bessel function of integer order. A separable $s$-wave
approximation, which dominates low energy scattering, is used through out this
work $V_{l}(k,q) \simeq \lambda_{l}k^{l}q^{l}$, namely  $V(\vc{k},\vc{q})
\simeq \lambda_{0}$. This approach is valid for  $k,q \ll 1/a$, $a$ being the
range of the potential. Therefore,  the action of the potential on the wave
function $V(r)\psi(\vc{r})$ is expressed in momentum space as
$(1/\Omega)\sum_{q}\lambda_{0}\psi(\vc{q})$ within this approximation.

\section{Scattering by a separable \texorpdfstring{$s$-wave}{s-wave} impurity}

\label{sec:AppImpScatt}

Here we solve the scattering problem for a single electron at very low
energies using the $s$-wave separable approximation. The obtained $T$-matrix
shows the leading behavior at low energies for short range impurities. We start
from~\eqref{eq:DiracRealSpace}, as written in momentum space and then transformed
to helicities. For a separable potential the result is
\begin{align}
i\,\partial_{t}\psi_{\sigma}(\vc{k},t)&=\sigma k\psi_{\sigma}(\vc{k},t)
+\frac{\lambda_{0}}{\Omega}
\left(
e^{i\theta_{k}\sigma^{1}/2}
\right)_{\sigma\sigma'}
\nonumber \\
&\times \sum_{q} \left(
e^{-i\theta_{q}\sigma^{1}/2}
\right)_{\sigma'\tau}\psi_{\tau}(\vc{q},t)\ .
\end{align}
The corresponding $T$-matrix equation will be written as
\begin{align}
\mat{T}(z;\vc{k}_{1},\vc{k}_{2})&=\lambda_{0}
\,e^{i\left(\theta_{k_1}-\theta_{k_2}\right)\sigma^{1}/2}\nonumber\\
&+\frac{1}{\Omega}\sum_{q}\mat{W}(\vc{k}_{1},\vc{q})
\mat{G}_{0}(z;\vc{q})\mat{T}(z;\vc{q},\vc{k}_{2})\ ,
\label{eq:TMatrix1P}
\end{align}
where $\mat{G}_{0}(z;\vc{q})$ is the diagonal propagator, whose  elements are
given by $1/(z-\sigma q)$.

The solution to~\eqref{eq:TMatrix1P} for an $s$-wave separable potential
can be sought in the form
$\mat{T}(z;\vc{k},\vc{q})=e^{i\theta_{k}\sigma^{1}/2}\mat{T}(z)\,
e^{-i\theta_{q}\sigma^{1}/2}$
where, after substitution of this ansatz in \eqref{eq:TMatrix1P} and
solving for $\mat{T}(z)$, we get the solution
$\mat{T}(z)  = \left[\lambda_{0}^{-1}-\mat{M}(z)\right]^{-1}$ with
$$
\mat{M}(z)  = \frac{1}{\left(2\pi\right)^{2}}
\int_{-\pi}^{\pi}\!\!d\theta e^{-i\theta\sigma^{1}/2}
\left[\int_{0}^{p_{c}}\!\!dk\, k\,\mat{G}_{0}(z;\vc{k})\right]
e^{i\theta\sigma^{1}/2}\ ,
$$
where $p_{c}$ is a cutoff on the order of the
smallest of the inverse of potential range $1/a$ or the inverse of
the lattice spacing $1/b$. When $|z|\ll p_{c},\,\Re(z)>0,\Im(z)>0$
the solution for the full $T$-matrix is found to be
\begin{equation}
\mat{T}(z;\vc{k}_{1},\vc{k}_{2})=
\frac{1}{\lambda_{0}^{-1}-\frac{z}{2\pi}\left[\ln\frac{z}{p_{c}}-i\frac{\pi}{2}
\right]} \left[\begin{array}{cc}
C_{12} & iS_{12}\\
iS_{12} & C_{12}
\end{array}\right]\ ,
\label{eq:TMatSolImpurityScatt}
\end{equation}
where for brevity we define $C_{12}=\cos(\theta_1/2-\theta_2/2)$
and $S_{12}=\sin(\theta_1/2-\theta_2/2)$

\section{Elliptic coordinates}

\label{sec:AppElliptic-coordinates}

In this appendix, the momentum $q=q_{x}+iq_{y}\in\mathbb{C}$ and we will write
explicitly $|q|$ for the modulus. The total momentum is, by convention, real and
positive, $Q>0$. The transformation to elliptic coordinates reads $q =
(Q/2)\cosh(u+iv)$ with $u\geq 0$ and $-\pi<v\leq\pi$ (see
Ref.~\onlinecite{spiegel1968schaum}).

All integrals needed in the main text are of the form
\begin{align}\label{eqC1}
\int_{0}^{u_{c}}du\int_{-\pi}^{\pi}&dv\frac{Q^{2}}{4}
|\sinh\left(u+iv\right)|^{2}e^{in_{+}\theta_{+}}e^{in_{-}\theta_{-}}\nonumber \\
&\times \left\{ \frac{1}{z\pm Q\cos(v)},\,\frac{1}{z\pm Q\cosh(u)}\right\}\ ,
\end{align}
where $n_{\pm}= 0,-1,1$ and
$$
e^{i\theta_{+}} = \frac{\cosh\left(\frac{u+iv}{2}\right)}
{\cosh\left(\frac{u-iv}{2}\right)}\ ,\quad
e^{i\theta_{-}} = -\frac{\sinh\left(\frac{u+iv}{2}\right)}
{\sinh\left(\frac{u-iv}{2}\right)}\ .
$$
The results after integration are collected in Table~\ref{tabIntegrals}. The
cutoffs in the different coordinate systems are chosen as $p_{c} = (Q/2)\sinh
u_{c}$, $u_{c}  \simeq \ln(4p_{c}/Q)$ and therefore
$(1/4)\sinh\left(2u_{c}\right) = 2p_{c}^{2}/Q^{2}+\mathcal{O}(1)$.

\begin{table*}[ht]
\begin{tabular}{|c|c|c|}
\hline
$\left(n_{+},\, n_{-}\right)$ & $\intop_{-\pi}^{\pi}dv\,|
 \sinh\left(u+iv\right)|^{2}e^{in_{+}\theta_{+}}e^{in_{-}\theta_{-}}$
& $\intop_{0}^{u_{c}}du\,|\sinh\left(u+iv\right)|^{2}e^{in_{+}
\theta_{+}}e^{in_{-}\theta_{-}}$\tabularnewline
\hline
\hline
$(0,0)$ & $\pi\left(2\sinh^{2}u+1\right)$
& $I_{s2}+u_{c}\sin^{2}v$\tabularnewline
\hline
$(0,\pm1)$ & $\pi\cosh u$
& $-I_{s2}\cos v+I_{c1}\sin^{2}v$\tabularnewline
\hline
$(\pm1,0)$ & $\pi\cosh u$
& $I_{s2}\cos v+I_{c1}\sin^{2}v$\tabularnewline
\hline
$(\pm1,\pm1)$ & $\pi$
& $-I_{c2}\cos\left(2v\right)+u_{c}\cos^{2}v$\tabularnewline
\hline
$(\pm1,\mp1)$ & $-\pi\left(2\sinh^{2}u-1\right)$
& $-I_{s2}+u_{c}\sin^{2}v$\tabularnewline
\hline
\end{tabular}
\caption{Table of integrals for the evaluation of Eq.\ \eqref{eqC1}. The imaginary parts drop out from the second column
and are not shown in the third because they are odd in $v$. Here we have
defined
$I_{s2} = \int_{0}^{u_{c}}du\,\sinh^{2}u=(1/4)
 \left(\sinh\left(2u_{c}\right)-2u_{c}\right)
 \simeq 2p_{c}^{2}/Q^{2}-u_{c}/2$,
$I_{c2} = \int_{0}^{u_{c}}d u\,\cosh^{2}u=
 (1/4)\left(\sinh\left(2u_{c}\right)+2u_{c}\right)
 \simeq 2p_{c}^{2}/Q^{2}+u_{c}/2$, and
$I_{c1} = \int_{0}^{u_{c}}d u\,\cosh u=\sinh u_{c}=2p_{c}/Q$.
}\label{tabIntegrals}
\end{table*}

Using Table~\ref{tabIntegrals}, we find the integrals $a$,  $b$,  $c$,  $d$ of
the main text as follows
\newcommand{\x}{{\textstyle\frac{z}{Q}}}
\newcommand{\ofx}{{\bigl(\textstyle\frac{z}{Q}}\bigr)}
\begin{align}
a(z)
& = \frac{1}{16\pi Q h\ofx}\left[ 2p_{c}^{2} \rho\ofx +
 \frac{zQ}{2} g\ofx\right]\ , \nonumber \\
b(z)
& = \frac{1}{16\pi Q h\ofx}\left[2p_{c}^{2} \rho^{2}\!\ofx -
\frac{Q^{2}}{2}g\ofx\right]\ , \nonumber\\
c(z)
& = \frac{1}{16\pi Q h\ofx}\left[ 2p_{c}^{2}+\left(z^{2}-
\frac{3Q^{2}}{2}\right)g\ofx\right]\ ,\nonumber \\
d(z)
& = \frac{1}{16\pi Q h\ofx}\left[ 2p_{c}^{2}-\left(z^{2}-
\frac{Q^{2}}{2}\right)g\ofx\right]\ ,
\label{eq:MPrimeSolution2}
\end{align}
where $\rho(x)$, $h(x)$ and $g(x)$ are defined in the main text after
Eq.~\eqref{eq:tMatValuesX}. The leading term in the cutoff is proportional to
$p_c^2$. The function $g(z/Q)$ is a correction of logarithmic order.

The eigenvalues and eigenvectors of the matrix $\mat{M}$,  which consists of the
expressions given in~\eqref{eq:MPrimeSolution2}, are given in the main text
[before Eq.~\eqref{eq:MPrimeSpectralDecomp} and in
Eqs.~\eqref{eq:tMatValuesX}],  except for the eigenvectors
$v_{s\pm}^{\intercal}(z) = [a_{\pm}(z),\pm b_{\pm}(z),\pm
b_{\pm}(z),a_{\pm}(z)]$. The ratio of the entries reads
\begin{equation}
 \frac{a(z)_\pm}{b(z)_\pm}
= \frac{b(z)_\mp}{a(z)_\mp}
= \sqrt{1 + \frac{c(z) - b(z)}{4 |a(z)|}} \pm \frac{c(z) - b(z)}{4 |a(z)|}\ ,\label{eq:Eigenvector+-}
\end{equation}
where, according to Eqs.~\eqref{eq:MPrimeSolution2}, $c(z)-b(z)>0$ and $a(z)<0$
for $z>Q$.

\section{Projectors for \texorpdfstring{$s$, $s+$, and $s-$}{s, s+, and s-}}
In Eq.\ \eqref{eq:resonantT_s} of the main text, we have considered the $(++)$-$(++)$ component of the asymmetric projector matrix
$\left(\mat{U}^{\dagger}(q)
v_{a}v_{a}^{\intercal}
\mat{U}(q')\right)_{++,++} \equiv X_a(q,q')$, with the result
\begin{equation}\label{eq:aProjector}
X_a(q,q')
=\frac{1}{2}\sin\left(\frac{\theta_{-}-\theta_{+}}{2}\right)
\sin\left(\frac{\theta'_{-}-\theta'_{+}}{2}\right) .
\end{equation}
Close to the resonances (see Sec.\ \ref{subsecFeshbachResonances}), we need also the symmetric projectors.
There, we find
\begin{align}
X_s(q,q')
&=\frac{1}{2}\sin\left(\frac{\theta_{-}+\theta_{+}}{2}\right)
\sin\left(\frac{\theta'_{-} + \theta'_{+}}{2}\right) ,\\
X_{s\pm}(q,q')
&=\textstyle
\left[a_\pm(z) \cos\left(\frac{\theta_{-}+\theta_{+}}{2}\right)
  \mp b_\pm(z) \cos\left(\frac{\theta_{-}-\theta_{+}}{2}\right)\right] \nonumber \\
&\times \textstyle
\left[a_\pm(z) \cos\left(\frac{\theta_{-}'+\theta_{+}'}{2}\right)
  \mp b_\pm(z) \cos\left(\frac{\theta_{-}'-\theta_{+}'}{2}\right)\right], \nonumber
\end{align}
where $a_\pm(z)$ and $b_\pm(z)$ are assumed to be normalized such that $2(a_\pm^2+b_\pm^2) = 1$.
In the isotropic limit ($Q \to 0$) the expression
${a(z)_+}/{b(z)_+} = {b(z)_-}/{a(z)_-}$ in Eq.\ \eqref{eq:Eigenvector+-} tends to infinity.
So, we have $a_+ \to 1/\sqrt{2}$, $b_+ \to 0$, $a_- \to 0$, and $b_- \to 1/\sqrt{2}$, such that the above expression simplifies to
\begin{equation}
X_{s\pm}(q,q') \to \frac{1}{2}
\cos\left(\frac{\theta_{-} \pm \theta_{+} }{2}\right)
\cos\left(\frac{\theta_{-}'\pm \theta_{+}'}{2}\right) \ .
\end{equation}

\

\section{Isotropic limit \texorpdfstring{$Q \rightarrow 0$}{Q->0}}

\label{sec:AppIsotrLimit}

\paragraph{\texorpdfstring{$(++)$}{(++)} scattering channel.}
As discussed in the main text, the isotropic limit for this situation can be
obtained by $Q\rightarrow 0$ leaving $\epsilon \gg Q$ fixed. This limit is
straightforward from Eqs.\eqref{eq:tMatValuesX}. Rotational symmetry makes the
states $s,s-$ degenerate. The space spanned by $v_s, v_{s+}$ has $(1,0,0,0)$
and $(0,0,0,1)$ as basis vectors. The state $s-$ has a very simple spinor,
$v_{s-}=(0,1,1,0)/\sqrt{2}$, but it is projected to zero in the $(++)$ channel.

\paragraph{\texorpdfstring{$(+-),(-+)$}{(+-),(-+)} scattering channels.}
The limit we consider in this situation is with fixed incoming momenta, so
$\epsilon \sim Q \cos(\theta_i)$,  being $q_i=|q_i|e^{i \theta_i}$ the incoming
momentum. We here make a definite assumption of entrance in $(+-)$ and such that
$0\leq \theta_i \leq \pi/2$, in order to ease the notation. After a long but
straightforward computation, the dominant ($Q \rightarrow 0$) on-shell
scattering amplitude elements are found to be
\begin{eqnarray}
t_{a}  &=& \pm \frac{\lambda_0 Q^2}{2|q_i q_f|}\,\sin(q_i)\sin(q_f)\ ,
\nonumber  \\
t_{s}  &=&     -\frac{2\pi Q e^{i \theta_i}}{p_c^2}\,\sin(q_i)\sin(q_f)\ ,
\nonumber  \\
t_{s+} &=& \pm \frac{4\pi Q}{p_c^2 f_{+}(\theta_i)}\ , \nonumber  \\
t_{s-} &=&     \frac{4\pi Q}{p_c^2 f_{-}(\theta_i)}\cos(q_i)\cos(q_f)\ ,
\end{eqnarray}
with
\begin{eqnarray*}
f_{\pm}(\theta) & =& \cos(\theta)+i\big[ 2\csc(\theta)-\sin(\theta) \big]
                   \nonumber\\
                &  & \pm i \sqrt{3+\cos^2(\theta)} \csc(\theta)e^{-i \theta} \ ,
\end{eqnarray*}
where the sign $\pm$ is chosen according to $\cos(\theta_i)=\pm \cos(\theta_f)$.

\addcontentsline{toc}{section}{\refname}
\bibliography{Graphene2B}

\end{document}